\documentclass[prl, twocolumn, superscriptaddress]{revtex4-2}
\usepackage{bm, amsmath, amsfonts, amssymb, ascmac, mathtools, braket}
\usepackage{times}
\usepackage{multirow}
\usepackage{graphicx}
\usepackage{float, color, xcolor}

\usepackage{bbm}
\usepackage{tabularx}

\usepackage{comment}

\usepackage[
pagebackref=false,
colorlinks=true,
linkcolor=blue,
urlcolor=blue,
filecolor=black,
citecolor=red,
pdfstartview=FitV,
pdftitle={},
pdfauthor={},
pdfsubject={},
pdfkeywords={},
pdfpagemode=None,
bookmarksopen=true
]{hyperref}

\newcommand{\ii}{\text{i}}

\begin{document}

\title{Lieb-Schultz-Mattis Theorem in Open Quantum Systems}

\author{Kohei Kawabata}
\email{kawabata@issp.u-tokyo.ac.jp}
\affiliation{Department of Physics, Princeton University, Princeton, New Jersey 08544, USA}
\affiliation{Institute for Solid State Physics, University of Tokyo, Kashiwa, Chiba 277-8581, Japan}

\author{Ramanjit Sohal}
\affiliation{Department of Physics, Princeton University, Princeton, New Jersey 08544, USA}

\author{Shinsei Ryu}
\affiliation{Department of Physics, Princeton University, Princeton, New Jersey 08544, USA}

\date{\today}

\begin{abstract}
The Lieb-Schultz-Mattis (LSM) theorem provides a general constraint on quantum many-body systems and plays a significant role in the Haldane gap phenomena and topological phases of matter.
Here, we extend the LSM theorem to open quantum systems and establish a general theorem that restricts the steady state and spectral gap of Liouvillians based solely on symmetry.
Specifically, we demonstrate that the unique gapped steady state is prohibited when translation invariance and $\mathrm{U} \left( 1 \right)$ symmetry are simultaneously present for noninteger filling numbers.
As an illustrative example, we find that no dissipative gap is open in the spin-$1/2$ dissipative Heisenberg model while a dissipative gap can be open in the spin-$1$ counterpart---an analog of the Haldane gap phenomena in open quantum systems.
Furthermore, we show that the LSM constraint manifests itself in a quantum anomaly of the dissipative form factor of Liouvillians.
We also find the LSM constraints due to symmetry intrinsic to open quantum systems, such as Kubo-Martin-Schwinger symmetry.
Our work leads to a unified understanding of phases and phenomena in open quantum systems.
\end{abstract}

\maketitle

Spectral gaps are pivotal for characterization of quantum phases of matter~\cite{Sachdev-textbook, Fradkin-textbook}.
While it is generally nontrivial to find out the ground state and spectral gap in quantum many-body systems, 
universal ingredients such as symmetry can provide their general understanding. 
The Lieb-Schultz-Mattis (LSM) theorem~\cite{LSM-61} is a prime example, 
which generally shows that the symmetry-preserving unique gapped ground state is prohibited for noninteger filling numbers in the simultaneous presence of translation invariance and $\mathrm{U} \left( 1 \right)$ symmetry~\cite{Affleck-86, Oshikawa-97, *Yamanaka-97}.
This theorem resolves a part of the Haldane conjecture~\cite{Haldane-83PLA, *Haldane-83PRL, Affleck-89, Haldane-17RMP, Tasaki-textbook}, prohibiting the gapped ground state in quantum spin chains with half-integer spins $S=1/2, 3/2, \cdots$.
For integer spins $S=0, 1, 2, \cdots$, on the other hand, no such general constraints are imposed, and a gap can be open---Haldane gap.
The LSM theorem has been further generalized to higher dimensions~\cite{Oshikawa-00, Hastings-04, Nachtergaele-07, Bachmann-20}, other symmetries~\cite{Chen-11, Parameswaran-13, Watanabe-15, *Watanabe-16, Ogata-19, *Ogata-21, Yao-20, *Yao-21, Tasaki-22}, and fermionic systems~\cite{Hsieh-16, Cheng-19, Aksoy-21}.
The LSM constraint is also a manifestation of a quantum anomaly in condensed-matter systems~\cite{Furuya-17, Cheng-16, Cho-17, Metlitski-18, Else-20, Gioia-22, Cheng-22}.
However, it has remained largely unexplored whether the LSM constraint emerges even in the presence of dissipation.

Spectral gaps are also crucial for open quantum systems.
As a result of coupling to the external environment, open quantum systems are no longer described by Hamiltonians,
but instead by Liouvillians that act on density operators~\cite{Nielsen-textbook, Breuer-textbook, Rivas-textbook}.
In general, Liouvillians are non-Hermitian superoperators and possess the complex-valued spectra.
The dissipative gap between the steady state and the first decaying state provides a timescale of the relaxation process and is fundamental for the open quantum dynamics~\cite{Cai-13, Znidaric-15, Medvedyeva-16, Shibata-19-Kitaev, *Shibata-19-AshkinTeller, Mori-20, Yoshida-23}.
The role of symmetry and topology in open quantum systems has also attracted growing interest~\cite{Diehl-11, Buca-12, Albert-14, Bardyn-18, Minganti-18, Coser-19, Lieu-20, Tonielli-20, *Huang-22, McGinley-20, Altland-21, deGroot-22, Sa-22, Kawabata-22}.
Despite the significance and recent interest,
the dissipative gap was investigated mainly for specific models.
Accordingly, general theories on the steady state and dissipative gap in open quantum systems, akin to the LSM theorem in closed quantum systems, have yet to be established.
Such a general theoretical understanding should be relevant to the control of quantum materials and further exploration of quantum technology.

In this Letter, we present a general theorem that restricts the spectral gaps of open quantum systems solely based on symmetry, generalizing the LSM theorem.
In particular, we demonstrate that the unique gapped steady state is prohibited in the simultaneous presence of translation invariance and $\mathrm{U} \left( 1 \right)$ symmetry for noninteger filling numbers.
As an illustrative example, we find that no dissipative gap is open in the $S=1/2$ dissipative Heisenberg model while a dissipative gap can be open in the $S=1$ counterpart---an open quantum analog of the Haldane gap phenomena.
As a unique feature with no analogs in closed quantum systems, our analysis elucidates a fundamental difference between strong and weak symmetries in open quantum systems.
We also find the LSM constraint arising from symmetry inherent to open quantum systems, Kubo-Martin-Schwinger (KMS) symmetry.

\textit{Gap and symmetry}.---We consider generic open quantum systems described by the Lindblad master equation $d\rho/dt = \mathcal{L} \rho$ with the Lindbladian~\cite{GKS-76, Lindblad-76, Nielsen-textbook, Breuer-textbook, Rivas-textbook}
\begin{align}
    \mathcal{L} \rho &= -\ii \left[ H, \rho \right] + \sum_{n} \left[ L_n \rho L_n^{\dag} - \frac{1}{2} \left\{ L_n^{\dag} L_n, \rho \right\} \right],
        \label{eq: Lindblad}
\end{align}
where $H$ is a Hermitian Hamiltonian that describes the unitary dynamics, and $L_{n}$'s are dissipators that describe the nonunitary coupling to the environment.
To investigate the Lindbladian spectrum, it is useful to double the Hilbert space and map $\mathcal{L}$ and $\rho$ to an operator and a state, respectively.
In particular, we map the density operator $\rho = \sum_{ij} \rho_{ij} \ket{i} \bra{j}$ to a pure state $\ket{\rho} = \sum_{ij} \rho_{ij} \ket{i} \ket{j}$ in the double Hilbert space.
Through this operator-state mapping, the Lindblad equation reduces to $d\ket{\rho}/dt  = \mathcal{L} \ket{\rho}$ with 
\begin{align}
    &\mathcal{L} = -\ii H_{+} + \ii H_{-} \nonumber \\
    &\quad + \sum_{n} \left[ L_{n, +} L_{n, -}^{*} - \frac{1}{2}\,( L_{n, +}^{\dag} L_{n, +} + L_{n, -}^{T} L_{n, -}^{*} ) \right],
        \label{eq: boson operator-state mapping}
\end{align}
where $O_{\pm}$ denotes an operator acting on the ket and bra spaces, respectively, defined from an operator $O$ acting on the original Hilbert space.
While Eq.~(\ref{eq: boson operator-state mapping}) assumes bosonic systems, the operator-state mapping can be similarly carried out for fermionic systems~\cite{Schmutz-78, Dzhioev-11, *Dzhioev-12, Kawabata-22}.

Through the operator-state mapping, we calculate the Lindbladian spectrum by diagonalizing the non-Hermitian many-body operator $\mathcal{L}$.
The real part of its eigenvalues is constrained to be nonpositive because of the contractive nature of the Lindblad dynamics.
The steady state corresponds to the zero eigenvalue of $\mathcal{L}$.
We define the dissipative gap $\Delta$ as the negative real part of the second largest eigenvalue (i.e., eigenvalue of the first decaying state).
This can be considered as a many-body generalization of line gap~\cite{Esaki-11, Shen-18, Kawabata-19}.
The dissipative gap $\Delta$ gives the relaxation timescale toward the steady state.
The Lindbladian is gapped (gapless) if $\Delta$ is nonvanishing (vanishing) in the infinite-size limit $V \to \infty$, and the gapped (gapless) Lindbladian is subject to the exponential (algebraic) relaxation process~\cite{Cai-13}.
The power-law behavior of the dissipative gap also yields the dynamical critical exponent.

We derive a general theorem that constrains the dissipative gap solely by symmetry.
First, we assume lattice translation invariance of the Lindbladian, $\mathcal{T} \mathcal{L} \mathcal{T}^{-1} = \mathcal{L}$, with the lattice translation operator $\mathcal{T}$.
Additionally, we assume $\mathrm{U} \left( 1 \right)$ symmetry in the individual ket and bra spaces, $U_{\pm} \mathcal{L} U_{\pm}^{-1} = \mathcal{L}$, which yields a conserved charge $N_{\pm}$.
We focus on the steady-state subspace with $N_{+} = N_{-}$ and define the filling number $\nu \coloneqq N_{\pm}/V$.
The $\mathrm{U} \left( 1 \right)$ charge typically corresponds to the total magnetization in spin systems and the total particle number in electron systems.
We can also introduce $\mathrm{U} \left( 1 \right)$ symmetry in the total Hilbert space, which is referred to as weak symmetry in contrast to strong symmetry defined above~\cite{Buca-12, Albert-14, Sa-22, Kawabata-22}.
We later show the different roles of strong and weak symmetries in the LSM constraints.

\textit{Lieb-Schultz-Mattis theorem}.---Now, we demonstrate the theorem:

\smallskip
{\bf Theorem.} In open quantum systems with lattice translation invariance and strong $\mathrm{U} \left( 1 \right)$ symmetry, if 
the Lindbladian is gapped and exhibits a unique steady state
in the subspace with the fixed $\mathrm{U} \left( 1 \right)$ charge, the filling number $\nu$ is required to be an integer.
In other words, if $\nu$ is not an integer,
the Lindbladian is gapless or exhibits degenerate steady states.

\smallskip
We prove this theorem, following Oshikawa's argument~\cite{Oshikawa-00}.
We consider a generic Lindbladian in $d$ dimensions with symmetry, where the system length in each direction is denoted by $L_i$ ($i=1, 2, \cdots, d$).
From $\mathrm{U} \left( 1 \right)$ symmetry, we introduce the $\mathrm{U} \left( 1 \right)$ flux $\phi_{\pm}$ in the individual ket or bra space.
Such a $\mathrm{U} \left( 1 \right)$ flux can be added by the twist operator or twisted boundary conditions, and physically corresponds to a magnetic flux in electronic systems.
Similarly to Ref.~\cite{Oshikawa-00}, we assume that the dissipative gap remains nonvanishing in the presence of the $\mathrm{U} \left( 1 \right)$ flux if it is originally open.
While this assumption is needed but has yet to be rigorously justified even for the original LSM theorem~\cite{Watanabe-18}, it is expected to hold since the thermodynamic quantities including the spectral gaps should not crucially change just by twisting the boundary conditions.
Under this assumption, if the dissipative gap is closed in the course of the insertion of the $\mathrm{U} \left( 1 \right)$ flux, the Lindbladian is gapless.

Then, let us consider a steady state $\ket{\rho_0}$ in the subspace with the fixed $\mathrm{U} \left( 1 \right)$ charge $N_{\pm}$ and the filling number $\nu = N_{\pm}/V$, and insert the $\mathrm{U} \left( 1 \right)$ flux $\phi_{+}$ in the ket space perpendicular to the $i=1$ direction~\cite{left}.
In the course of the adiabatic insertion of the unit flux $\phi_{+} = 2\pi$, the Lindbladian spectrum flows and goes back to the original spectrum.
The original steady state $\ket{\rho_0}$ changes to another eigenstate $U_{+}^{\rm (twist)} \ket{\rho_0}$ of the Lindbladian $\mathcal{L}$ with the twist operator $U_{+}^{\rm (twist)} \coloneqq e^{2\pi\ii \sum_{j=1}^{L_1} jn_{j, +}/L_1}$, where $n_{j, +}$ is the local density of the $\mathrm{U} \left( 1 \right)$ charge $N_{+} = \sum_{j=1}^{L_1} n_{j, +}$.
Importantly, the twist operator evolves by translation as
\begin{align}
    \mathcal{T} U_{+}^{\rm (twist)} \mathcal{T}^{-1} = U_{+}^{\rm (twist)} e^{-2\pi\ii N_{+}/L_1}.
\end{align}
Hence, when $e^{-\ii P_{0}}$ is the eigenvalue of translation $\mathcal{T}$ for the steady state $\ket{\rho_0}$ (i.e., $\mathcal{T} \ket{\rho_0} = e^{-\ii P_{0}} \ket{\rho_0}$), we have 
\begin{align}
    \mathcal{T}\,(U_{+}^{\rm (twist)} \ket{\rho_0}) = e^{-\ii\,(P_{0} + 2\pi N_{+}/L_1)} \,(U_{+}^{\rm (twist)} \ket{\rho_0}),
\end{align}
meaning that $U_{+}^{\rm (twist)} \ket{\rho_0}$ is an eigenstate of $\mathcal{T}$ with the eigenvalue $e^{-\ii\,(P_{0} + 2\pi N_{+}/L_1)}$ and orthogonal to the original steady state $\ket{\rho_0}$ unless $N_{+}/L_1 = \nu \sum_{i=2}^{d} L_{i}$ is an integer~\cite{coprime}.
Thus, for noninteger $\nu$, we can always have another steady state $U_{+}^{\rm (twist)} \ket{\rho_0}$ with the different eigenvalue of $\mathcal{T}$, which implies the gapless Lindblad spectrum or the degeneracy of the steady states.
In other words, the gapped Lindbladian with the unique steady state requires the integer filling number $\nu \in \mathbb{Z}$.

While we here focus on Lindbladians, this theorem is applicable to more generic non-Hermitian operators.
Additionally, translation invariance is required for the above discussions.
This implies that the conclusion of the LSM theorem can be violated under the open boundary conditions or in open quantum systems subject to dissipation only at boundaries.
Furthermore, the $\mathrm{U} \left( 1 \right)$ flux $\phi_{\pm}$ in the individual ket or bra space breaks invariance under modular conjugation, which is required for physical Lindbladians~\cite{Kawabata-22}.
Still, the adiabatic insertion of $\phi_{\pm}$ can detect the ingappability of open quantum systems.

\begin{figure}[t]
\centering
\includegraphics[width=\linewidth]{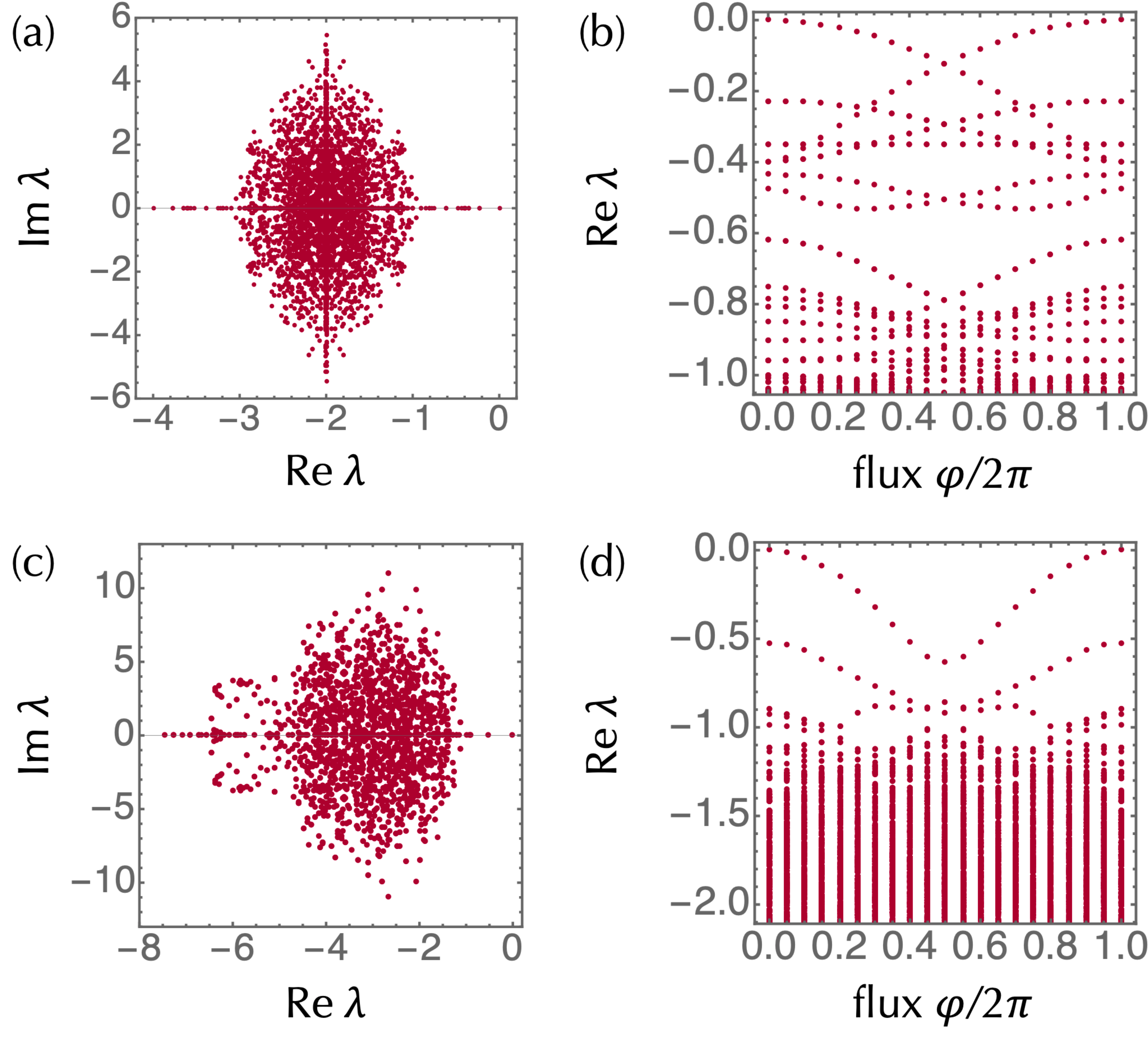} 
\caption{Dissipative Heisenberg XXZ model ($J=J_z=1.0$, $\gamma = 1.0$) for (a, b)~spin half $S=1/2$ ($L=8$) and (c, d)~spin one $S=1$ ($L=5$).
The $\mathrm{U} \left( 1 \right)$ charge is $S^{z}_{\pm} = 0$, i.e., half filling (a, b)~$\nu = 1/2$ for $S=1/2$ and (c, d)~$\nu = 1$ for $S=1$.
The $\mathrm{U} \left( 1 \right)$ flux $\phi$ in the ket space is inserted.
(a, c)~Lindbladian spectrum with $\phi = 0$ for (a)~$S=1/2$ and (c)~$S=1$.
(b, d)~Real part of the Lindbladian spectrum around the steady state $\lambda = 0$ for (b)~$S=1/2$ and (d)~$S=1$ as a function of the flux $\phi$.}
	\label{fig: Lindblad-Heisenberg}
\end{figure}

\textit{Dissipative Heisenberg XXZ models}.---As an illustrative example that shows the significance of the LSM theorem, we study the dissipative Heisenberg XXZ model in one dimension,
\begin{align}
    H &= \sum_{n=1}^{L} \left[ J \left( S_{n}^{x} S_{n+1}^{x} + S_{n}^{y} S_{n+1}^{y} \right) + J_{z} S_{n}^{z} S_{n+1}^{z} \right], \label{eq: XXZ - Ham} \\
    L_{n} &= \sqrt{\gamma} S_{n}^{z} \quad \left( n = 1, 2, \cdots, L \right), \label{eq: XXZ - diss}
\end{align}
where $S_{n}^{i}$'s ($i=x, y, z$) are quantum spin operators of spin number $S$ at site $n$.
The dissipators $L_{n}$'s describe the dephasing process of spin coherence~\cite{Breuer-textbook}.
The Lindbladian $\mathcal{L}$ respects $\mathrm{U} \left( 1 \right)$ symmetry in the individual ket and bra spaces,
\begin{align}
    [\mathcal{L}, S_{\pm}^{z}] = 0,\quad
    S_{\pm}^{z} \coloneqq \sum_{n=1}^{L} S_{n, \pm}^{z},
\end{align}
and conserves the total magnetization $S_{\pm}^{z}$ along the $z$ axis.
Hence, it is subject to the LSM constraint and cannot exhibit the unique gapped steady state except for the integer filling number $\nu = S_{\pm}^{z}/L+S \in \mathbb{Z}$.
In particular, for the half filling $S_{\pm}^{z} = 0$, we have the half-integer filling number for half-integer spins $S = 1/2, 3/2, \cdots$ and the integer filling number for integer spins $S = 0, 1, 2, \cdots$.
Consequently, the LSM theorem prohibits the unique gapped steady state for half-integer spins whereas no such constraints are imposed for integer spins, akin to the Haldane gap phenomena of the antiferromagnetic Heisenberg model~\cite{Haldane-83PLA, *Haldane-83PRL, Affleck-89, Haldane-17RMP, Tasaki-textbook}.

Using the operator-state mapping in Eq.~(\ref{eq: boson operator-state mapping}), we study the Lindbladian spectra for $S=1/2$ and $S=1$ (Fig.~\ref{fig: Lindblad-Heisenberg})~\cite{symmetry}.
For the half-integer spin $S=1/2$, the dissipative gap closes in the course of the insertion of the $\mathrm{U} \left( 1 \right)$ flux [Fig.~\ref{fig: Lindblad-Heisenberg}\,(b)].
Here, the $\mathrm{U} \left( 1 \right)$ fluxes $\phi_{\pm}$ in the ket and bra spaces are introduced by imposing the twisted boundary conditions, 
\begin{align}
    S_{n+L, \pm}^{+} = e^{\ii \phi_{\pm}} S_{n, \pm}^{+},\quad
    S_{n+L, \pm}^{-} = e^{-\ii \phi_{\pm}} S_{n, \pm}^{-}
        \label{eq: twisted boundary conditions}
\end{align}
with the spin raising and lowering operators, $S_{n, \pm}^{+} \coloneqq S_{n, \pm}^{x} + \ii S_{n, \pm}^{y}$ and $S_{n, \pm}^{-} \coloneqq S_{n, \pm}^{x} - \ii S_{n, \pm}^{y}$.
The gap closing due to the flux insertion implies the gapless spectrum, which is compatible with the LSM theorem~\cite{gap}.

For the integer spin $S=1$, by contrast, the dissipative gap remains open even in the course of the insertion of the $\mathrm{U} \left( 1 \right)$ flux [Fig.~\ref{fig: Lindblad-Heisenberg}\,(d)], which is prohibited for $S=1/2$.
The distinct spectral properties between half-integer and integer spins are reminiscent of the Haldane gap phenomena and show the significant predictability of the LSM theorem.
The influence of the $\mathrm{U} \left( 1 \right)$ flux is merely twisting the boundary conditions and hence intuitively expected to become less significant when we increase the system size $L$.
Thus, Fig.~\ref{fig: Lindblad-Heisenberg}\,(d) may imply a nonzero gap even in the infinite-size limit $L \to \infty$.
This is in a similar spirit to the assumption made in Oshikawa’s argument~\cite{Oshikawa-00}.
However, the gap opening in Fig.~\ref{fig: Lindblad-Heisenberg}\,(d) may be due to a finite-size effect that should be studied carefully in future work.
Even if the aforementioned intuitive argument works in closed quantum systems, it may break down in open quantum systems.

The gapped ground state of the $S=1$ antiferromagnetic Heisenberg model further exhibits the symmetry-protected topological phase~\cite{AKLT-87, *AKLT-88, Kennedy-Tasaki-92, Gu-09, *Chen-13, Pollmann-10, *Pollmann-12}.
By contrast, the steady state of the dissipative $S=1$ Heisenberg model is given as the identity (i.e., infinite-temperature state) and does not exhibit nontrivial topological properties.
In other dissipative spin models, the steady state can exhibit topological phases.
The LSM theorem does not necessarily enforce the unique gapped steady state for $S=1$;
a gapless steady state can appear for different dissipators even for $S=1$~\cite{supplement}.
Away from the integer filling $\nu \neq 1$, the gap opening is prohibited even for $S=1$.
In the absence of $\mathrm{U} \left( 1 \right)$ symmetry, the LSM theorem is inapplicable, and the dissipative gap can be open even for $S=1/2$.
In other words, if we are to open the dissipative gap for $S=1/2$ without degenerate steady states, we need to break $\mathrm{U} \left( 1 \right)$ symmetry or translation invariance.
Consistently, the dissipative Ising model without $\mathrm{U} \left( 1 \right)$ symmetry was shown to exhibit the unique gapped steady state~\cite{Cai-13, Znidaric-15}.
Importantly, the LSM theorem is not only applicable to bosonic systems but also fermionic systems~\cite{supplement}.

\textit{Quantum anomaly}.---The original LSM theorem is related to a quantum anomaly in closed systems~\cite{Furuya-17, Cheng-16, Cho-17, Metlitski-18, Else-20, Gioia-22, Cheng-22}.
Similarly, we find that the LSM theorem developed in this Letter is a manifestation of a quantum anomaly in open systems.
For clarity, we study generic one-dimensional Lindbladians.
Let us first introduce the $\mathrm{U} \left( 1 \right)$ fluxes $\phi_{\pm}$ by twisting the boundary conditions as in Eq.~(\ref{eq: twisted boundary conditions}).
Because of the twisted boundary conditions, translation invariance $\mathcal{T}$ is broken.
However, the twisted Lindbladian is invariant under the generalized translation, $\mathcal{T} \left( \phi_{\pm} \right) \mathcal{L} \left( \phi_{\pm} \right) \mathcal{T}^{-1} \left( \phi_{\pm} \right) = \mathcal{L} \left( \phi_{\pm} \right)$, for $\mathcal{T} \left( \phi_{\pm} \right) \coloneqq e^{\ii \phi_{+} (n_{1,+}-\nu_+)} e^{\ii \phi_{-} (n_{1,-}-\nu_-)} \mathcal{T}$ with the local density $n_{j, \pm}$ of the strong $\mathrm{U} \left( 1 \right)$ charge (i.e., $N_{\pm} = \sum_{j=1}^{L} n_{j, \pm}$). 
This twisted translation operator satisfies $\mathcal{T} \left( \phi_{+} + 2\pi, \phi_{-} \right) =  e^{- 2\pi \ii \nu_+} \mathcal{T} \left( \phi_{+}, \phi_{-} \right)$ and
\begin{align}
 \mathcal{T}^{L} \left( \phi_{\pm} \right) = e^{\ii \phi_{+} (N_{+}-L\nu_+)} e^{\ii \phi_{-} (N_{-}-L\nu_-)}. 
    \label{eq: T^L relation}
\end{align}

While the choice of a partition function is nontrivial in open quantum systems, we study the twisted dissipative form factor,
\begin{align}
    &Z \left( T, L, \phi_{\pm}, \psi_{\pm}, l \right) \coloneqq \mathrm{tr} \left[  e^{T \mathcal{L}\,(\phi_{\pm})} e^{\ii \psi_{+} N_{+}} e^{\ii \psi_{-} N_{-}} \mathcal{T}^{l} \left( \phi_{\pm} \right) \right]
\end{align}
with $T, \psi_{\pm} \in \mathbb{R}$.
This is considered as the form factor of the real-time open quantum dynamics and should capture the dissipative gap~\cite{Can-19, Anish-22}.
To characterize the anomaly in a fixed filling sector of strong $\mathrm{U} \left( 1 \right)$ symmetry, we Fourier transform $Z \left( T, L, \phi_{\pm}, \psi_{\pm}, l \right)$ and obtain the projected dissipative form factor,
\begin{align}
	&\tilde{Z}_{q_{\pm}} \left( T, L, \phi_{\pm}, l \right) \coloneqq \nonumber \\
    &\quad \int_{0}^{2\pi} \frac{d\psi_+ d\psi_-}{\left( 2\pi \right)^2} e^{-\ii q_{+} \psi_{+} - \ii q_{-} \psi_{-}} Z \left( T, L, \phi_{\pm}, \psi_{\pm}, l \right).
\end{align}
Then, we have
\begin{align}
	& \tilde{Z}_{q_{\pm}} \left( T, L, \phi_{+} + 2\pi, \phi_{-}, l  \right) = e^{-2\pi \ii l \nu_+} \tilde{Z}_{q_{\pm}} \left( T, L, \phi_{+}, \phi_{-}, l \right), 
    \label{eq: u1 flux insertion anomaly}
\end{align}
and
\begin{align}
	& \tilde{Z}_{q_{\pm}} \left( T, L, \phi_{\pm}, l + L \right) \nonumber \\
    &\quad = e^{\ii\phi_+ (q_+ - L\nu_+) + \ii \phi_- (q_- - L\nu_-)} \tilde{Z}_{q_{\pm}} \left( T, L, \phi_{\pm}, l \right)
    \label{eq: translation anomaly}
\end{align}
from Eq.~\eqref{eq: T^L relation}.
Thus, for the noninteger filling $\nu_{\pm} \notin \mathbb{Z}$, the projected dissipative form factor $\tilde{Z}_{q_{\pm}} \left( T, L, \phi_{\pm}, l \right)$ acquires the nontrivial quantum phases, signaling a mixed anomaly between the strong $\mathrm{U} \left( 1 \right)$ symmetry and weak translation symmetry---LSM anomaly.
For the integer filling $\nu_{\pm} \in \mathbb{Z}$, on the other hand, such nontrivial phases do not appear, consistent with no LSM constraints.
Notably, our discussion is also applicable to discrete quantum channels $\mathcal{E}$, for which the untwisted dissipative form factor can be introduced as $\mathrm{tr}\,\mathcal{E}^{m}$.

In contrast to strong $\mathrm{U} \left( 1 \right)$ symmetry, no anomaly arises for weak $\mathrm{U} \left( 1 \right)$ symmetry, $[\mathcal{L}, N] = 0$, with the weak $\mathrm{U} \left( 1 \right)$ charge $N \coloneqq N_+ - N_-$ in the total Hilbert space. 
Using the twisted translation operator $\mathcal{T} \left( \phi \right) \coloneqq e^{\ii \phi\,(n_{1,+} - n_{1,-} - \nu)} \mathcal{T}$, we introduce the twisted dissipative form factor 
$Z \left( T, L, \phi, \psi, l \right) \coloneqq \mathrm{tr} \left[ e^{T \mathcal{L}\,(\phi)} e^{\ii \psi N}  \mathcal{T}^{l} \left( \phi \right) \right]$ and the projected one $\tilde{Z}_{q} \left( T, L, \phi, l \right) \coloneqq \int \left( d\psi/2\pi \right) e^{-\ii q \psi} Z \left( T, L, \phi, \psi, l \right)$. 
In the steady-state subspace with $N=0$ and $q=0$, we have $\tilde{Z}_{0} \left( T, L, \phi + 2\pi, l  \right) = \tilde{Z}_{0} \left( T, L, \phi, l + L \right) = \tilde{Z}_{0} \left( T, L, \phi, l \right)$, implying the absence of quantum anomaly. 
Consistently, gapless excitation cannot be detected by twisting Lindbladians in both ket and bra spaces~\cite{supplement}.
These discussions show an important difference between strong and weak symmetries in open quantum systems.

\textit{Kubo-Martin-Schwinger symmetry}.---The LSM theorem can be generalized to other symmetry including discrete symmetry in a similar manner to closed quantum systems~\cite{Chen-11, Parameswaran-13, Watanabe-15, *Watanabe-16, Ogata-19, *Ogata-21, Yao-20, *Yao-21, Tasaki-22, Hsieh-16, Cheng-19, Aksoy-21}.
For example, the unique gapped steady state is prohibited also in dissipative quantum spin models with $\mathbb{Z}_2 \times \mathbb{Z}_2$ spin-flip symmetry~\cite{supplement}.

Notably, KMS symmetry---symmetry inherent in thermal equilibrium~\cite{Kubo-57, Martin-Schwinger-59, Sieberer-15, Glorioso-lecture, Liu-Ryu-21}---also yields the LSM constraint.
For illustration, let us consider a translation-invariant Lindbladian $\mathcal{L}$ in one dimension that consists of Majorana fermions $\lambda_{n}$'s ($n=1, 2, \cdots, L$), satisfying $\{ \lambda_{m}, \lambda_{n} \} = 2\delta_{mn}$.
We assume that the system length $L$ is even.
Because of the Hermiticity-preserving nature, $\mathcal{L}$ is generally invariant under modular conjugation $\mathcal{J}$, defined by an antiunitary operator $\mathcal{J}$ satisfying~\cite{Ojima-80, Witten-18-review, Kawabata-22}
\begin{align}
    \mathcal{J} \lambda_{n, \pm} \mathcal{J}^{-1} = \lambda_{n, \mp}.
\end{align}
To introduce KMS symmetry, we also consider another antiunitary operation
\begin{align}
    \mathcal{R} \lambda_{n, \pm} \mathcal{R}^{-1} = \lambda_{n, \pm}
\end{align}
with an antiunitary operator $\mathcal{R}$.
If the Lindbladian $\mathcal{L}$ is invariant under $\mathcal{R}$, it is also invariant under the combination of $\mathcal{R}$ and $\mathcal{J}$.
Thus, we introduce KMS symmetry $U_{\rm KMS} \mathcal{L} U_{\rm KMS}^{-1} = \mathcal{L}$ by the unitary operator $U_{\rm KMS} \coloneqq \mathcal{J}\mathcal{R}$ with $U_{\rm KMS}^2 = 1$.
In contrast to ordinary symmetry, KMS symmetry $U_{\rm KMS}$ accompanies the exchanges of the ket and bra degrees of freedom.

We find that the interplay of KMS symmetry $U_{\rm KMS}$ and translation invariance $\mathcal{T}$ also leads to the LSM constraint.
The key is the nontrivial algebra~\cite{supplement}
\begin{align}
    \mathcal{T} U_{\rm KMS} \mathcal{T}^{-1} = - U_{\rm KMS}.
        \label{eq: KMS}
\end{align}
Consequently, all the eigenvalues of the Lindbladian $\mathcal{L}$, including the zero eigenvalue of the steady states, are at least two-fold degenerate.
To see this, let $\ket{\rho}$ be an eigenstate of $\mathcal{L}$, and $\lambda$ and $k \in \{ +1, -1 \}$ be their eigenvalues (i.e., $\mathcal{L} \ket{\rho} = \lambda \ket{\rho}$, $U_{\rm KMS} \ket{\rho} = k \ket{\rho}$).
Owing to translation invariance $\mathcal{T}\mathcal{L}\mathcal{T}^{-1} = \mathcal{L}$ and Eq.~(\ref{eq: KMS}), we also have $\mathcal{L}\,(\mathcal{T}\ket{\rho}) = \lambda\,(\mathcal{T}\ket{\rho})$ and $U_{\rm KMS}\,(\mathcal{T}\ket{\rho}) = - k\,(\mathcal{T}\ket{\rho})$, which implies that $\mathcal{T}\ket{\rho}$ is another eigenstate of $\mathcal{L}$ that belongs to the same eigenvalue $\lambda$ but has the different eigenvalue $-k$ of $U_{\rm KMS}$, i.e., degeneracy of the Lindbladian spectrum.
This is the LSM constraint in Majorana Lindbladians, an open quantum analog of that in Majorana Hamiltonians~\cite{Hsieh-16, Cheng-19, Aksoy-21}.
This LSM constraint makes the dissipative form factor vanish, which also signals a quantum anomaly.
Such degeneracy of the Lindbladian spectrum should affect the steady-state properties and dynamics of open quantum systems.

The combination of translation invariance and fermion parity symmetry also gives rise to the degeneracy of the Lindbladian spectrum.
Here, while strong fermion parity $\left( -1 \right)^{F_{\pm}} = \prod_{n=1}^{L/2} \left( \ii \lambda_{2n-1, \pm} \lambda_{2n, \pm} \right)$ in the individual ket or bra space yields such a LSM constraint, weak fermion parity $\left( -1 \right)^{\cal F} = \left( -1 \right)^{F_{+}} \left( -1 \right)^{F_{-}}$ in the total Hilbert space gives no such constraints.
Although KMS symmetry is also relevant to both ket and bra degrees of freedom, it still yields the LSM constraint, implying a unique role of KMS symmetry in open quantum systems.

\textit{Discussions}.---Spectral gaps are crucial for understanding closed and open quantum systems.
In this Letter, we establish the LSM theorem in open quantum systems, which provides a general constraint on their steady state and dissipative gap solely by symmetry.
As a consequence of the LSM constraint, we discover a fundamental distinction between half-integer and integer spins---an open quantum analog of the Haldane gap phenomena.
It merits further research to investigate the dissipative Haldane gap in various analytical and numerical approaches.
Since our discussions on quantum anomaly rely solely on symmetry and the structure of the double Hilbert space, they should also be relevant to non-Markovian Liouvillians, which we leave for future work.

The LSM theorem developed in this Letter gives a guiding principle to understand quantum phases of open systems.
Specifically, when a dissipative gap is open in the presence of symmetry for noninteger filling, the LSM theorem ensures the nontrivial degeneracy of the steady states, typically originating from spontaneous symmetry breaking or topological order.
In closed quantum systems, such LSM-type constraints also prohibit short-range-entangled states.
Similarly, our LSM theorem should prohibit short-range-entangled states in the double Hilbert space.
It merits further study to clarify its connection with entanglement properties of mixed states~\cite{Hastings-11}.
It should also be noted that the mere presence of a dissipative gap does not necessarily lead to short-range correlations of the steady state in contrast with closed quantum systems~\cite{Poulin-10}.

Our formalism based on the double Hilbert space may find applications to other physical systems, such as disordered systems~\cite{Kimchi-18}.
A quantum-channel formulation of average-symmetry-protected topological phases has recently been developed~\cite{Ma-22}.
It is worthwhile to further explore a  relationship between disorder and dissipation.
Moreover, a unique feature of open quantum systems is the non-Hermitian skin effect~\cite{Song-19, Haga-21, Liu-20PRR, Yang-22, Kawabata-23}, which makes the spectral properties under the open boundary conditions distinct~\cite{supplement}. 
The skin effect is captured by the complex-spectral winding~\cite{Fukui-98Nucl, Gong-18, Zhang-22, Kawabata-22MB} and the concomitant quantum anomaly~\cite{Kawabata-21TFt}, thereby having a potential connection with the LSM theorem.
The role of pseudospectra~\cite{Trefethen-Embree-textbook} in the LSM theorem is also worth studying.

\begingroup
\renewcommand{\addcontentsline}[3]{}
\begin{acknowledgments}
K.K. thanks Haruki Watanabe for his seminar that stimulates this work, and Ryohei Kobayashi, Yuhan Liu, and Ken Shiozaki for helpful discussions.
K.K. is supported by the Japan Society for the Promotion of Science (JSPS) through the Overseas Research Fellowship.
This work is supported by the National Science Foundation under Grant No.~DMR-2001181, Simons Investigator Grant from the Simons Foundation (Grant No.~566116), and the Gordon and Betty Moore Foundation through Grant No.~GBMF8685 toward the Princeton theory program.
\end{acknowledgments}
\endgroup

\let\oldaddcontentsline\addcontentsline
\renewcommand{\addcontentsline}[3]{}
\bibliography{LSM.bib}

\begin{thebibliography}{109}
\bibitem [{lef()}]{left}%
  \BibitemOpen
  \href@noop {} {}\bibinfo {note} {{The following discussion is also applicable
  to left eigenstates of Lindbladians. In passing, the identity is generally a
  left eigenstate of Lindbladians with zero eigenvalue.}}\BibitemShut {Stop}%
\bibitem [{cop()}]{coprime}%
  \BibitemOpen
  \href@noop {} {}\bibinfo {note} {{For $d \geq 2$ and $\nu = p/q$ with coprime
  integers $p$, $q$, we should choose $L_2, L_3, \cdots, L_d$ such that they
  are not divided by $q$.}}\BibitemShut {Stop}%
\bibitem [{sym()}]{symmetry}%
  \BibitemOpen
  \href@noop {} {}\bibinfo {note} {{The reflection symmetry of the Lindbladian
  spectrum for $S=1/2$ about $\mathrm{Re}\,\lambda = - \gamma L/4$ is due to
  antiunitary symmetry $\mathcal{C}_{\pm} \left( \mathcal{L} + \gamma L/4
  \right) \mathcal{C}^{-1}_{\pm} = - \left( \mathcal{L} + \gamma L/4 \right)$
  with $\mathcal{C}_{\pm} \coloneqq ( \prod_{n=1}^{L} S_{n, \pm}^{z}
  )\,\mathcal{K}$~\cite{Prosen-12, Kawasaki-22, Sa-22, Kawabata-22}. This
  symmetry is specific to $S=1/2$ and not respected for $S=1$.}}\BibitemShut
  {Stop}%
\bibitem [{gap()}]{gap}%
  \BibitemOpen
  \href@noop {} {}\bibinfo {note} {{The gap of the dissipative $S=1/2$ XXZ
  model was shown to behave $\Delta \propto 1/L^2$ under the open boundary
  conditions~\cite{Cai-13, Znidaric-15}.}}\BibitemShut {Stop}%
\end{thebibliography}
\let\addcontentsline\oldaddcontentsline

\clearpage
\widetext

\setcounter{secnumdepth}{3}

\renewcommand{\theequation}{S\arabic{equation}}
\renewcommand{\thefigure}{S\arabic{figure}}
\renewcommand{\thetable}{S\arabic{table}}
\setcounter{equation}{0}
\setcounter{figure}{0}
\setcounter{table}{0}
\setcounter{section}{0}
\setcounter{tocdepth}{0}

\numberwithin{equation}{section} 

\begin{center}
{\bf \large Supplemental Material for 
``Lieb-Schultz-Mattis Theorem in Open Quantum Systems"}
\end{center}


\section{Dissipative Heisenberg XXZ model}
    \label{asec: Lindblad XXZ}

\begin{figure}[H]
\centering
\includegraphics[width=\linewidth]{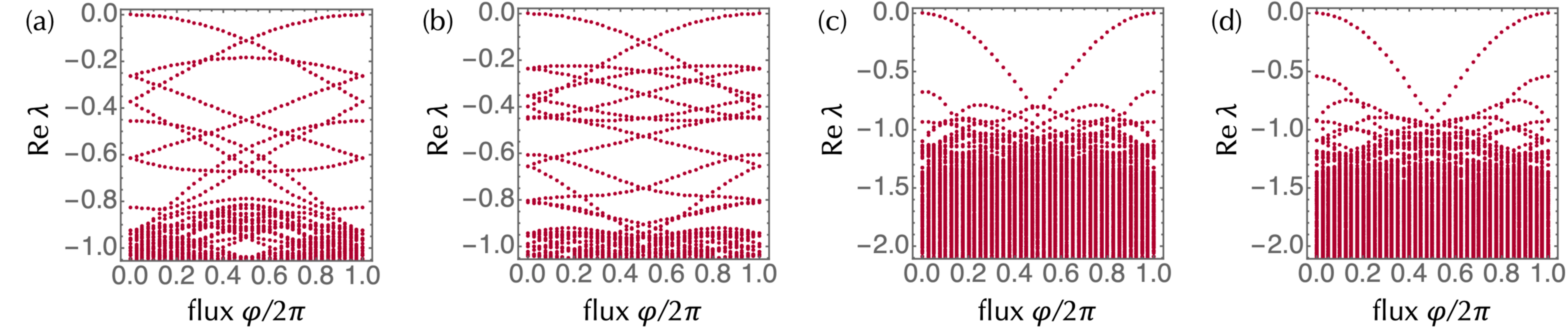} 
\caption{Complex-spectral flow of the dissipative Heisenberg XXZ model with dissipators $L_n = \sqrt{\gamma} S_{n}^{z}$ ($J = J_z = 1.0$, $\gamma = 1.0$) for (a, b)~spin half $S=1/2$ ($L=8$) and (c, d)~spin one $S=1$ ($L=5$).
The $\mathrm{U} \left( 1 \right)$ flux $\phi$ in the ket space is inserted.
The $\mathrm{U} \left( 1 \right)$ charge is (a)~$S_{\pm}^{z} = -2$ ($\nu = 1/4$), (b)~$S_{\pm}^{z} = -1$ ($\nu = 3/8$), (c)~$S_{\pm}^{z} = -2$ ($\nu = 3/5$), and (d)~$S_{\pm}^{z} = -1$ ($\nu = 4/5$).}
	\label{afig: Lindblad-Heisenberg-different-filling}
\end{figure}

In the main text, we study a dissipative Heisenberg XXZ model described by 
\begin{align}
    H &= \sum_{n=1}^{L} \left[ J \left( S_{n}^{x} S_{n+1}^{x} + S_{n}^{y} S_{n+1}^{y} \right) + J_{z} S_{n}^{z} S_{n+1}^{z} \right], 
        \label{aeq: XXZ} \\
    L_{n} &= \sqrt{\gamma}\,S_{n}^{z} \quad \left( n = 1, 2, \cdots, L \right),
        \label{aeq: XXZ - dissipator1} 
\end{align}
where $S_{n}^{i}$'s ($i=x, y, z$) are quantum spin operators of spin number $S$ at site $n$, and the periodic boundary conditions are imposed (i.e., $S_{n+L}^{i} = S_{n}^{i}$).
Through the operator-state mapping, the Lindbladian $\mathcal{L}$ reads
\begin{align}
    \mathcal{L} = & -\ii \sum_{n=1}^{L} \left[ J\,(S_{n, +}^{x} S_{n+1, +}^{x} + S_{n, +}^{y} S_{n+1, +}^{y}) + J_z\,S_{n, +}^{z} S_{n+1, +}^{z }\right] 
    + \ii \sum_{n=1}^{L} \left[ J\,(S_{n, -}^{x} S_{n+1, -}^{x} + S_{n, -}^{y} S_{n+1, -}^{y}) + J_z\,S_{n, -}^{z} S_{n+1, -}^{z}\right] \nonumber \\
    &\qquad\qquad\qquad\qquad\qquad + \gamma \sum_{n=1}^{L} \left[ S_{n, +}^{z} S_{n, -}^{z} - \frac{1}{2} ( ( S_{n, +}^{z})^2 + (S_{n, -}^{z})^2 ) \right],
\end{align}
where $S_{n, +}^{x, y, z}$'s and $S_{n, -}^{x, y, z}$'s are spin operators in the ket and bra spaces, respectively.
This Lindbladian respects $\mathrm{U} \left( 1 \right)$ symmetry in the individual ket and bra spaces,
\begin{align}
    [\mathcal{L}, S_{\pm}^{z}] = 0,\quad
    S_{\pm}^{z} \coloneqq \sum_{n=1}^{L} S_{n, \pm}^{z},
\end{align}
and conserves the total magnetization $S_{\pm}^{z}$ along the $z$ axis.
In contrast to Fig.~1 in the main text for the half filling, Fig.~\ref{afig: Lindblad-Heisenberg-different-filling} shows the complex-spectral flow of the dissipative Heisenberg XXZ model for the different filling.
For spin half $S=1/2$ [Fig.~\ref{afig: Lindblad-Heisenberg-different-filling}\,(a, b)], the dissipative gap closes in the course of the flux insertion in a similar manner to the half filling [see Fig.~1\,(b) in the main text].
On the other hand, for spin one $S=1$, we also find the gap closing due to the flux insertion for the noninteger filling numbers, which contrasts with the gap opening for the half filling [compare Fig.~\ref{afig: Lindblad-Heisenberg-different-filling}\,(c, d) with Fig.~1\,(d) in the main text].
These results are consistent with the Lieb-Schultz-Mattis (LSM) theorem developed in the main text.

\begin{figure}[H]
\centering
\includegraphics[width=0.5\linewidth]{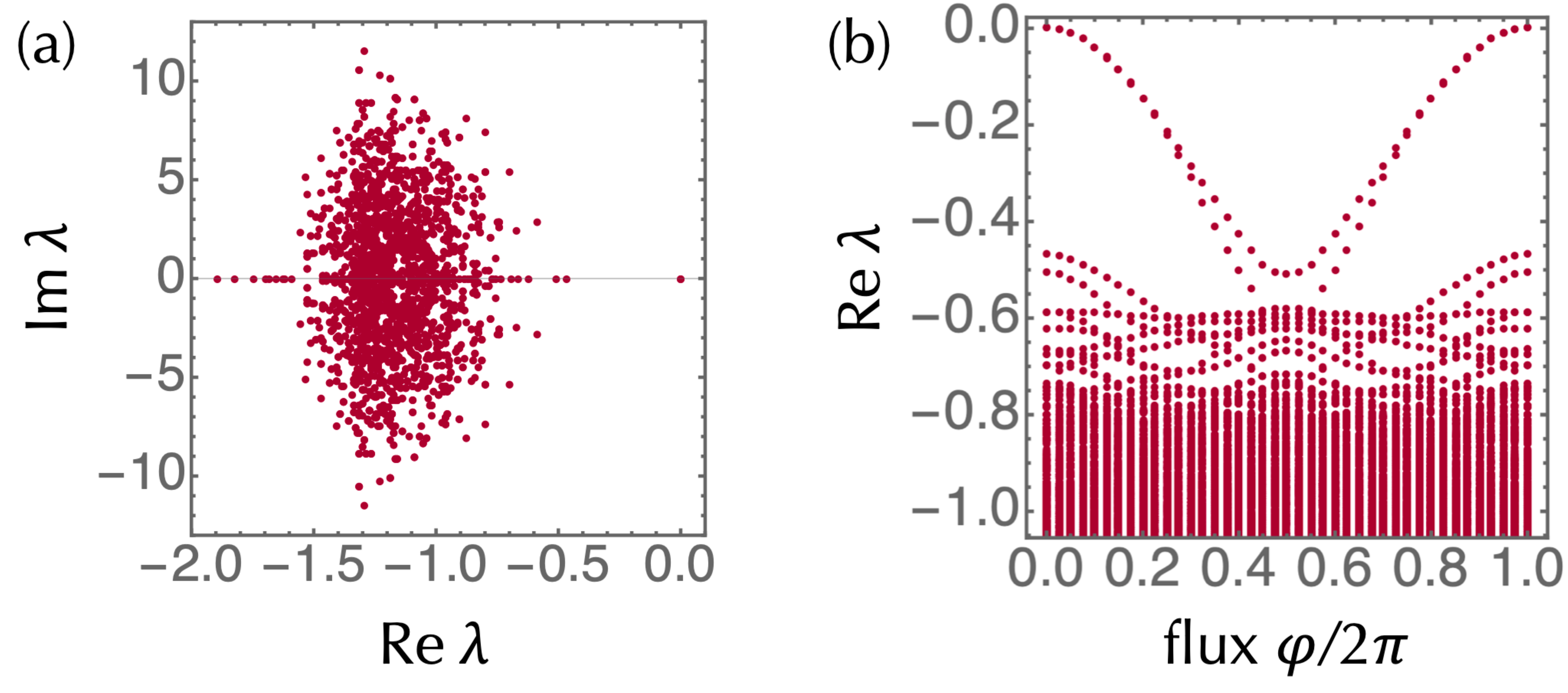} 
\caption{Dissipative Heisenberg XXZ model with dissipators $L_n = \sqrt{\gamma}\,(S_{n}^{z})^2$ and spin one $S=1$ ($L=5$, $J=J_z=1.0$, $\gamma = 1.0$).
The $\mathrm{U} \left( 1 \right)$ charge is $S^{z}_{\pm} = 0$, i.e., half filling $\nu = 0$.
The $\mathrm{U} \left( 1 \right)$ flux $\phi$ in the ket space is inserted.
(a)~Lindbladian spectrum with $\phi = 0$.
(b)~Real part of the Lindbladian spectrum around the steady state $\lambda = 0$ as a function of the flux $\phi$.}
	\label{afig: Lindblad-Heisenberg-spin1}
\end{figure}

As another dissipative spin model, we investigate a dissipative Heisenberg XXZ model described by Eq.~(\ref{aeq: XXZ}) and 
\begin{align}
    L_{n} &= \sqrt{\gamma}\,(S_{n}^{z})^2 \quad \left( n = 1, 2, \cdots, L \right).
        \label{aeq: XXZ - dissipator2} 
\end{align}
Through the operator-state mapping, the Lindbladian $\mathcal{L}$ reads
\begin{align}
    \mathcal{L} = & -\ii \sum_{n=1}^{L} \left[ J\,(S_{n, +}^{x} S_{n+1, +}^{x} + S_{n, +}^{y} S_{n+1, +}^{y}) + J_z\,S_{n, +}^{z} S_{n+1, +}^{z }\right] 
    + \ii \sum_{n=1}^{L} \left[ J\,(S_{n, -}^{x} S_{n+1, -}^{x} + S_{n, -}^{y} S_{n+1, -}^{y}) + J_z\,S_{n, -}^{z} S_{n+1, -}^{z}\right] \nonumber \\
    &\qquad\qquad\qquad\qquad\qquad + \gamma \sum_{n=1}^{L} \left[ (S_{n, +}^{z})^2 (S_{n, -}^{z})^2 - \frac{1}{2} ( (S_{n, +}^{z})^4 + (S_{n, -}^{z})^4 ) \right].
\end{align}
We study the complex spectrum of this dissipative Heisenberg XXZ model for $S=1$ under the boundary conditions twisted by $\mathrm{U} \left( 1 \right)$ symmetry (Fig.~\ref{afig: Lindblad-Heisenberg-spin1}):
\begin{align}
    S_{n+L, \pm}^{+} = e^{\ii \phi_{\pm}} S_{n, \pm}^{+},\quad
    S_{n+L, \pm}^{-} = e^{-\ii \phi_{\pm}} S_{n, \pm}^{-}.
\end{align}
Notably, in the absence of the $\mathrm{U} \left( 1 \right)$ flux, this Lindbladian exhibits the two-fold degenerate steady state in the subspace with the fixed $\mathrm{U} \left( 1 \right)$ charge.
One of the steady states remains gapped in the insertion of the $\mathrm{U} \left( 1 \right)$ flux. 
By contrast, the other is absorbed into the excitation spectrum, implying the gapless nature.

\section{Dissipative fermionic model}
    \label{asec: Liouvillian skin}

\begin{figure}[H]
\centering
\includegraphics[width=0.5\linewidth]{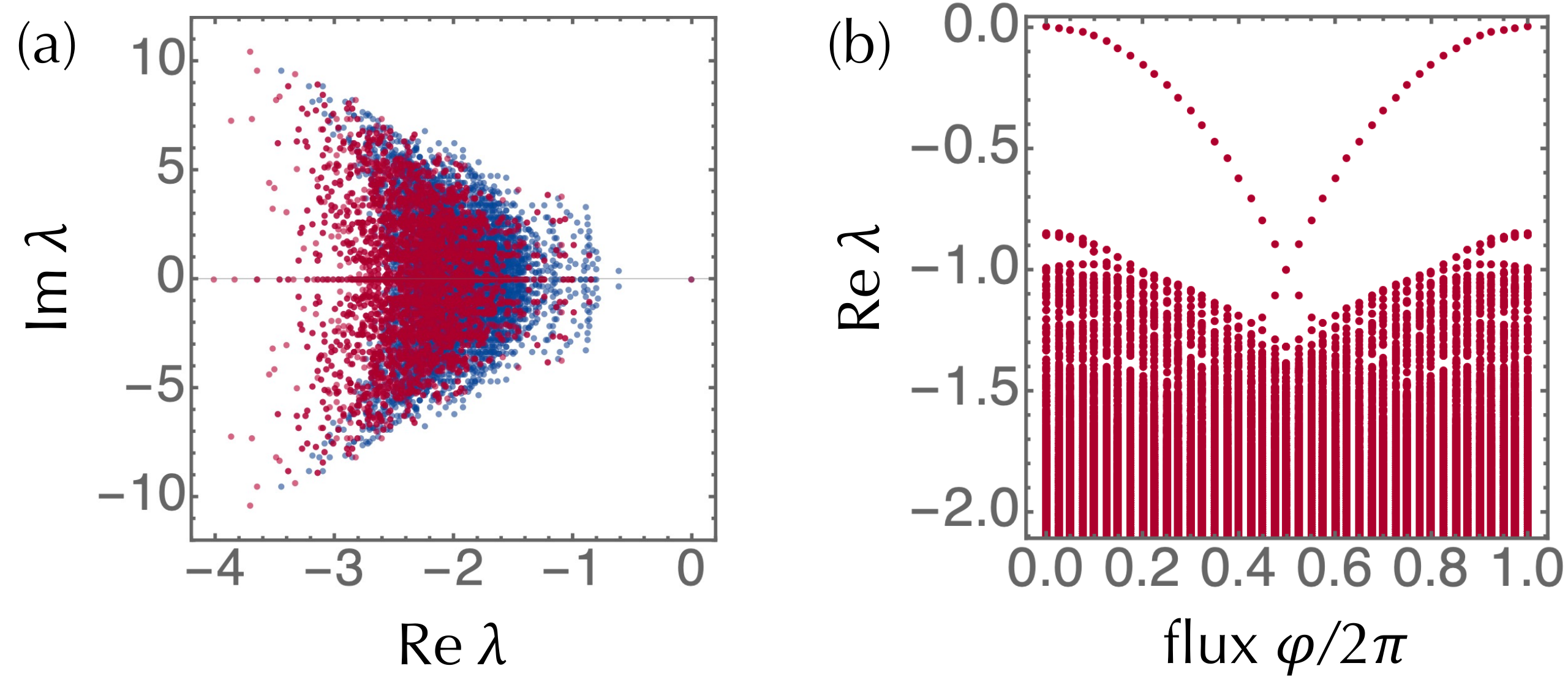} 
\caption{Dissipative fermionic model ($L=8$, $J=1.0$, $\gamma = 1.0$) with the half filling $\nu = 1/2$.
The $\mathrm{U} \left( 1 \right)$ flux $\phi$ in the ket space is inserted.
(a)~Lindbladian spectrum under the periodic boundary conditions with $\phi = 0$ (red dots) and open boundary conditions (blue dots).
(b)~Real part of the Lindbladian spectrum around the steady state $\lambda = 0$ as a function of the flux $\phi$.}
	\label{fig: Lindblad-fermion}
\end{figure}

We study an open quantum fermionic system in one dimension described by the Lindblad master equation.
The Hamiltonian reads
\begin{align}
    H = J \sum_{n=1}^{L} \left( c_{n+1}^{\dag} c_{n} + c_{n}^{\dag} c_{n+1} \right),
        \label{aeq: fermion Hamiltonian}
\end{align}
where $c_{n}$ ($c_{n}^{\dag}$) annihilates (creates) a spinless fermion at site $n$, and $J \in \mathbb{R}$ denotes the coherent hopping amplitude.
The dissipators are given as the incoherent asymmetric hopping
\begin{align}
    L_{n} = \sqrt{\gamma} c_{n+1}^{\dag} c_{n} \quad \left( n=1, 2, \cdots, L \right),
        \label{aeq: fermion dissipator}
\end{align}
which pushes fermions from the left to the right with the strength $\gamma \geq 0$~\cite{Znidaric-15, Haga-21, Kawabata-23}.
Through the fermionic operator-state mapping~\cite{Schmutz-78, Dzhioev-11, *Dzhioev-12, Kawabata-22}, the Lindbladian reads
\begin{align}
    &\mathcal{L} = -\ii J \sum_{n=1}^{L} \left( c_{n+1, +}^{\dag} c_{n, +} + c_{n, +}^{\dag} c_{n+1, +} \right)
    +\ii J \sum_{n=1}^{L} \left( c_{n+1, -}^{\dag} c_{n, -} + c_{n, -}^{\dag} c_{n+1, -} \right) \nonumber \\
    &\qquad\qquad + \gamma \sum_{n=1}^{L} \left( c_{n+1, +}^{\dag} c_{n, +} c_{n+1, -}^{\dag} c_{n, -} - \frac{1}{2} n_{n, +} \left( 1- n_{n+1, +} \right) - \frac{1}{2} n_{n, -} \left( 1- n_{n+1, -} \right) \right),
\end{align}
where $c_{n, +}$ ($c_{n, -}$) denotes the spinless fermionic operator at site $n$ in the ket (bra) space, and $n_{n, \pm} \coloneqq c_{n, \pm}^{\dag} c_{n, \pm}$ denotes the particle number operator at site $n$.
This Lindbladian respects $\mathrm{U} \left( 1 \right)$ symmetry in the individual space, 
\begin{align}
    [\mathcal{L}, N_{\pm}] = 0,\qquad
    N_{\pm} \coloneqq \sum_{n=1}^{L} n_{n, \pm} = \sum_{n=1}^{L} c_{n, \pm}^{\dag} c_{n, \pm},
\end{align}
and conserves the total particle number in each ket or bra space.
Correspondingly, the twisted Lindbladian reads
\begin{align}
    &\mathcal{L} \left( \phi_{+}, \phi_{-} \right) = -\ii J \sum_{n=1}^{L} \left( e^{\ii \phi_{+}/L} c_{n+1, +}^{\dag} c_{n, +} + e^{- \ii \phi_{+}/L} c_{n, +}^{\dag} c_{n+1, +} \right)
    +\ii J \sum_{n=1}^{L} \left( e^{\ii \phi_{-}/L} c_{n+1, -}^{\dag} c_{n, -} + e^{-\ii \phi_{-}/L} c_{n, -}^{\dag} c_{n+1, -} \right) \nonumber \\
    &\qquad\qquad\qquad\qquad + \gamma \sum_{n=1}^{L} \left( e^{\ii\,(\phi_{+}+\phi_{-})/L} c_{n+1, +}^{\dag} c_{n, +} c_{n+1, -}^{\dag} c_{n, -} - \frac{1}{2} n_{n, +} \left( 1- n_{n+1, +} \right) - \frac{1}{2} n_{n, -} \left( 1- n_{n+1, -} \right) \right),
\end{align}
where the $\mathrm{U} \left( 1 \right)$ fluxes $\phi_{+}$ and $\phi_{-}$ in the ket and bra spaces are introduced uniformly.
Owing to the simultaneous presence of translation invariance and $\mathrm{U} \left( 1 \right)$ symmetry, the Lindbladian is subject to the 
LSM
constraint, which is consistent with the complex-spectral flow in Fig.~\ref{fig: Lindblad-fermion}\,(b).

A unique feature of this Lindbladian is the non-Hermitian skin effect~\cite{Song-19, Haga-21, Liu-20PRR, Yang-22, Kawabata-23}.
Its complex spectrum under the open boundary conditions is distinct from that under the periodic boundary conditions [Fig.~\ref{fig: Lindblad-fermion}\,(a)].
Notably, the skin effect occurs also in the single-particle Hilbert space.
Let $\ket{m, n} \coloneqq c_{m, +}^{\dag} c_{n, -}^{\dag} \ket{\rm vac}$ be the single-particle state at sites $m$ and $n$ in the double Hilbert space.
For clarity, we here consider the dissipation-only dynamics, i.e., $J = 0$.
Then, under the periodic boundary conditions, we have
\begin{align}
    \mathcal{L} \ket{n, n} = \gamma \left( \ket{n+1, n+1} - \ket{n, n} \right)
        \label{aeq: Liouvillian skin PBC}
\end{align}
for $n = 1, 2, \cdots, L$.
Notably, Eq.~(\ref{aeq: Liouvillian skin PBC}) is formulated in the subspace spanned solely by the diagonal states $\{ \ket{1, 1}, \ket{2, 2}, \cdots, \ket{L, L} \}$.
The other eigenstates superposed by off-diagonal states $\ket{m, n}$ ($m \neq n$) in the double Hilbert space do not contribute to the steady state or the dissipative gap.
Thanks to translation invariance, Eq.~(\ref{aeq: Liouvillian skin PBC}) is readily solved by the Fourier transform.
The steady state corresponds to the zero mode with zero momentum, and the dissipative gap is obtained as
\begin{align}
    \Delta = \gamma\,\mathrm{Re} \left( 1- e^{2\pi\ii/L} \right)
    = \gamma \left( 1 - \cos \frac{2\pi}{L} \right)
    \simeq \frac{2\pi^2 \gamma}{L^2} \quad \left( L \to \infty \right).
\end{align}
Thus, the Lindbladian is gapless, which is compatible with the LSM theorem.
Under the open boundary conditions, on the other hand, we have Eq.~(\ref{aeq: Liouvillian skin PBC}) for $n=1, 2, \cdots, L-1$ and 
\begin{align}
    \mathcal{L} \ket{L, L} = 0. 
\end{align}
As a consequence of the different boundary conditions, the Lindbladian spectrum is distinct, i.e., non-Hermitian skin effect.
Except for the zero eigenvalue corresponding to the steady state, all the eigenvalues of the Lindbladian coalesce into $-\gamma$.
Thus, the dissipative gap is obtained as
\begin{align}
    \Delta = \gamma,
\end{align}
which does not vanish even for $L \to \infty$.
Importantly, under the open boundary conditions, translation invariance is broken, and hence the LSM theorem is inapplicable.
The gap opening in the above Lindbladian implies an important role of the boundary conditions and the non-Hermitian skin effect in open quantum systems.
In this respect, it is also notable that the skin effect is captured by the complex-spectral winding~\cite{Fukui-98Nucl, Gong-18, Zhang-22, Kawabata-22MB}.
Previously, the generalized boundary conditions for which both boundary coupling amplitude and phase are arbitrary were shown to be useful for understanding the skin effect and the bulk-boundary correspondence in non-Hermitian systems~\cite{Xiong-18, Kunst-18, *Koch-20, CHLee-19}.
Since such generalized boundary conditions cannot be interpreted as the boundary conditions twisted by $\mathrm{U} \left( 1 \right)$ symmetry, the LSM-type discussions in this work are not straightforwardly applicable.

\section{Weak $\mathrm{U} \left( 1 \right)$ symmetry}
    \label{asec: twist}

\begin{figure}[H]
\centering
\includegraphics[width=\linewidth]{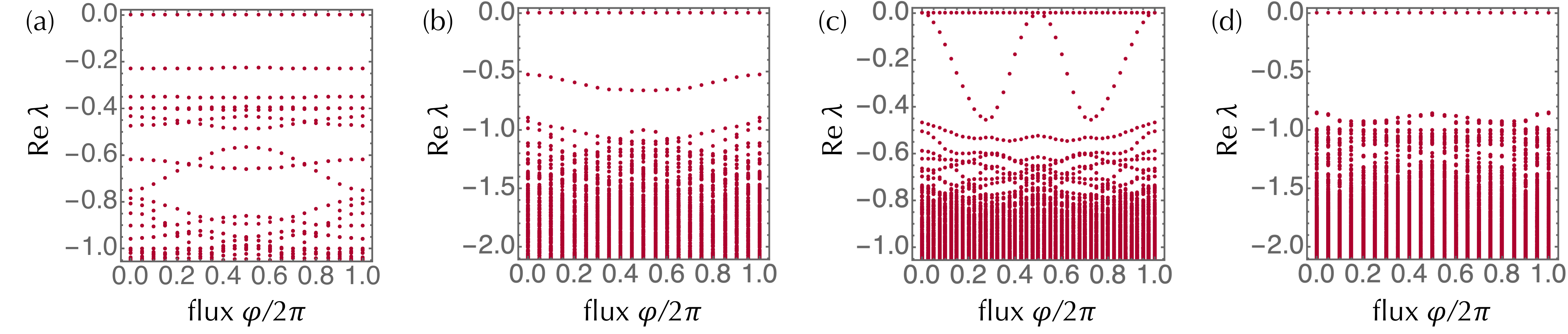} 
\caption{Complex-spectral flow of Lindbladians in the half filling [(a, d)~$\nu = 1/2$ and (b, c)~$\nu = 1$].
The $\mathrm{U} \left( 1 \right)$ fluxes $\phi_{+}$ and $\phi_{-}$ are inserted in both ket and bra spaces so that the Lindbladians will be invariant under modular conjugation (i.e., $\phi_{+} + \phi_{-} = 0$).
(a, b, c)~Dissipative Heisenberg XXZ model with (a)~dissipators $L_{n} = \sqrt{\gamma}\,S_{n}^{z}$ and spin $S=1/2$ ($L=8$, $J=J_z=1.0$, $\gamma = 1.0$), 
(b)~dissipators $L_{n} = \sqrt{\gamma}\,S_{n}^{z}$ and spin $S=1$ ($L=5$, $J=J_z=1.0$, $\gamma = 1.0$), 
and 
(c)~$L_{n} = \sqrt{\gamma}\,(S_{n}^{z})^2$ and spin $S=1$ ($L=5$, $J=J_z=1.0$, $\gamma = 1.0$).
(d)~Dissipative fermionic model with dissipators $L_{n} = \sqrt{\gamma}\,c_{n+1}^{\dag} c_{n}$ ($L=8$, $J=1.0$, $\gamma = 1.0$).
}
	\label{afig: Lindblad-braket}
\end{figure}

In Fig.~\ref{afig: Lindblad-braket}, we provide further results on the complex-spectral flow of the dissipative Heisenberg XXZ models in Eqs.~(\ref{aeq: XXZ}), (\ref{aeq: XXZ - dissipator1}), and (\ref{aeq: XXZ - dissipator2}) and the dissipative fermionic model in Eqs.~(\ref{aeq: fermion Hamiltonian}) and (\ref{aeq: fermion dissipator}).
Here, the $\mathrm{U} \left( 1 \right)$ fluxes $\phi_{+}$ and $\phi_{-} = - \phi_{+}$ are inserted in both ket and bra spaces so that the Lindbladian will be invariant under modular conjugation, $\mathcal{J} \mathcal{L} \mathcal{J}^{-1} = \mathcal{L}$, with an antiunitary operation $\mathcal{J}$ that exchanges the ket and bra degrees of freedom~\cite{Kawabata-22}.
As a consequence of modular conjugation symmetry, the steady state remains to have zero eigenvalue even in the presence of the $\mathrm{U} \left( 1 \right)$ fluxes.
In contrast with the $\mathrm{U} \left( 1 \right)$ twist only in the ket or bra space, the dissipative gap remains open for all the cases.
This behavior is consistent with the absence of quantum anomaly for weak $\mathrm{U} \left( 1 \right)$ symmetry in the total Hilbert space, as discussed in the main text.

\section{Lieb-Schultz-Mattis theorem for discrete symmetry}
    \label{asec: discrete}

We discuss the LSM theorem in open quantum systems for discrete symmetry.
For clarity, let us focus on dissipative quantum spin systems in one dimension.
We impose translation invariance, as well as $\mathbb{Z}_2 \times \mathbb{Z}_2$ spin-flip symmetry in the individual ket or bra space:
\begin{align}
    P_{\pm}^{i} \mathcal{L}\,( P_{\pm}^{i} )^{-1} = \mathcal{L},\quad
    P_{\pm}^{i} \coloneqq \exp \left( -\ii \pi \sum_{n=1}^{L} S_{n, \pm}^{i} \right)\quad
    \left( i = x, z \right).
\end{align}
Examples include the XYZ model $H = \sum_{n=1}^{L} \left( J_x S_{n}^{x} S_{n+1}^{x} + J_y S_{n}^{y} S_{n+1}^{y} + J_{z} S_{n}^{z} S_{n+1}^{z} \right)$ with the dissipators $L_{n}^{i} = \sqrt{\gamma^{i}} S_{n}^{i} S_{n+1}^{i}$ ($i=x, y, z$).
In a similar manner to Ref.~\cite{Yao-21} for closed quantum systems, we twist the boundary conditions by the symmetry generator $P_{\pm}^{z}$:
\begin{align}
    S^{i}_{n+L, \pm} = P_{\pm}^{z} S^{i}_{n, \pm}\,(P_{\pm}^{z})^{-1}
\end{align}
for $n = 1, 2, \cdots, L$.
The Lindbladian $\tilde{\cal L}$ with this $\mathbb{Z}_2$-symmetry-twisted boundary conditions remains to respect $\mathbb{Z}_2 \times \mathbb{Z}_2$ symmetry but is no longer invariant under translation.
However, if we introduce the twisted translation operation $\tilde{\cal T}_{\pm}$ by
\begin{align}
    \tilde{\cal T}_{\pm} \coloneqq P_{\pm}^{z} \mathcal{T},
\end{align}
the twisted Lindbladian $\tilde{\cal L}$ is invariant under $\tilde{\cal T}_{\pm}$.
Then, we have the nontrivial algebra
\begin{align}
    \tilde{\cal T}_{\pm} P_{\pm}^{x} \tilde{\cal T}^{-1}_{\pm} = \left( -1 \right)^{2S} P_{\pm}^{x}
\end{align}
with the spin number $S$.
Therefore, the nontrivial phase $\left( -1 \right)^{2S} = -1$ appears for half-integer spins $S=1/2, 3/2, \cdots$ while no such nontrivial phases appear for integer spins $S=0, 1, 2, \cdots$.
As a consequence of this nontrivial phase for half-integer spins $S=1/2, 3/2, \cdots$, all the eigenstates of the twisted Lindbladian $\tilde{\cal L}$, especially its steady states, are at least two-fold degenerate.
To see this, let $\ket{\rho}$ be an eigenstate of $\tilde{\cal L}$ and $\tilde{\cal T}_{\pm}$ with the eigenvalues $\lambda$ and $e^{-\ii P}$ (i.e., $\tilde{\cal L} \ket{\rho} = \lambda \ket{\rho}$, $\tilde{\cal T}_{\pm} \ket{\rho} = e^{-\ii P} \ket{\rho}$).
Then, we have
\begin{align}
    \tilde{\cal L}\,(P_{\pm}^{x}\ket{\rho}) = \lambda\,(P_{\pm}^{x}\ket{\rho}),\quad
    \tilde{\cal T}_{\pm}\,(P_{\pm}^{x}\ket{\rho}) = - e^{-\ii P}\,(P_{\pm}^{x}\ket{\rho}),
\end{align}
implying that $\tilde{\cal T}_{\pm}\ket{\rho}$ is another eigenstate of $\tilde{\cal L}$ that belongs to the same eigenvalue $\lambda$ but has the different eigenvalue $- e^{-\ii P} \neq e^{-\ii P}$ of $\tilde{\cal T}$.
If we further assume that the dissipative gap does not depend on the $\mathbb{Z}_2$-symmetry twist of the boundary conditions, similarly to Ref.~\cite{Yao-21} for closed quantum systems, the unique gapped steady state is prohibited.

This LSM constraint due to $\mathbb{Z}_2 \times \mathbb{Z}_2$ symmetry also manifests itself in a quantum anomaly of the dissipative form factor in a similar manner to that due to $\mathrm{U} \left( 1 \right)$ symmetry discussed in the main text.
Similarly to Ref.~\cite{Cheng-22}, we consider the twisted partition function
\begin{align}
    Z \left( T, L, k_{+}, k_{-}, m_{+}, m_{-}, n_{+}, n_{-} \right) \coloneqq \mathrm{tr} \left[ e^{T \tilde{\cal L}} \left( P_{+}^{z} \right)^{k_{+}} \left( P_{-}^{z} \right)^{k_{-}} \left( P_{+}^{x} \right)^{m_{+}} \left( P_{-}^{x} \right)^{m_{-}} \tilde{\cal T}_{+}^{n_{+}} \tilde{\cal T}_{-}^{n_{-}} \right]
\end{align}
with $k_{+}, k_{-}, m_{+}, m_{-}, n_{+}, n_{-} \in \mathbb{Z}$.
From the cyclic property of the trace, we have
\begin{align}
    Z \left( T, L, k_{+}, k_{-}, m_{+}, m_{-}, n_{+}, n_{-} \right) = \left( -1 \right)^{m_{\pm}} Z \left( T, L, k_{+}, k_{-}, m_{+}, m_{-}, n_{+}, n_{-} \right)
\end{align}
and hence
\begin{align}
    Z \left( T, L, k_{+}, k_{-}, m_{+} = 1, m_{-}, n_{+}, n_{-} \right) = Z \left( T, L, k_{+}, k_{-}, m_{+}, m_{-} = 1, n_{+}, n_{-} \right) = 0,
\end{align}
which signals a quantum anomaly.

\section{Lieb-Schultz-Mattis theorem for Kubo-Martin-Schwinger symmetry}
    \label{asec: KMS}

Let us study Majorana fermions $\lambda_{n, \pm}$'s ($n=1,2,\cdots, L$) on a one-dimensional chain in the double Hilbert space.
We normalize $\lambda_{n}$'s by $\{ \lambda_{m}, \lambda_{n} \} = 2\delta_{mn}$. 
The system length $L$ is assumed to be even.
We choose the operator basis such that $\lambda_{n, \pm}$'s are real for odd $n$ and pure imaginary for even $n$:
\begin{align}
    \mathcal{K} \lambda_{n, \pm} \mathcal{K}^{-1} = \left( -1 \right)^{n-1} \lambda_{n, \pm}.
\end{align}
We also introduce modular conjugation
\begin{align}
    \mathcal{J} \lambda_{n, \pm} \mathcal{J}^{-1} = \lambda_{n, \mp},
\end{align}
where $\mathcal{J}$ is an antiunitary operator constructed as~\cite{Kawabata-22}
\begin{align}   
    \mathcal{J} = \prod_{n=1}^{L/2} \frac{\lambda_{2n-1, +} - \lambda_{2n-1, -}}{\sqrt{2}} \frac{\ii\lambda_{2n, +} + \ii\lambda_{2n, -}}{\sqrt{2}} \mathcal{K}.
\end{align}
Because of their Hermiticity-preserving nature, arbitrary Lindbladians are invariant under modular conjugation.
We also introduce another type of modular conjugation by 
\begin{align}
    \tilde{\mathcal{J}} \lambda_{n, \pm} \tilde{\mathcal{J}}^{-1} = -\lambda_{n, \mp},
\end{align}
with 
\begin{align}   
    \tilde{\cal J} = \prod_{n=1}^{L/2} \frac{\lambda_{2n-1, +} + \lambda_{2n-1, -}}{\sqrt{2}} \frac{\ii\lambda_{2n, +} - \ii\lambda_{2n, -}}{\sqrt{2}} \mathcal{K}.
\end{align}
These two types of modular conjugation are related by
\begin{align}
    {\cal J} \tilde{\cal J} = \left( -1 \right)^{L/2} \left( -1 \right)^{\cal F},
\end{align}
with fermion parity in the total Hilbert space
\begin{align}
    \left( -1 \right)^{\cal F} \lambda_{n, \pm} \left( -1 \right)^{\cal F} = - \lambda_{n, \pm},\quad
    \left( -1 \right)^{\cal F} = \prod_{n=1}^{L/2} \left( \ii \lambda_{2n-1, +} \lambda_{2n, +} \right) \left( \ii \lambda_{2n-1, -} \lambda_{2n, -} \right).
        \label{aeq: fermion parity total}
\end{align}

Now, let us consider Kubo-Martin-Schwinger (KMS) symmetry with the unitary operations $\mathcal{J}\mathcal{K}$ and $\tilde{\cal J} \mathcal{K}$, satisfying 
\begin{align}
    (\mathcal{J}\mathcal{K})\,\lambda_{n, \pm}\,(\mathcal{J}\mathcal{K})^{-1} = \left( -1 \right)^{n-1} \lambda_{n, \mp},\quad
    (\tilde{\cal J} \mathcal{K})\,\lambda_{n, \pm}\,(\tilde{\cal J} \mathcal{K})^{-1} &= \left( -1 \right)^{n} \lambda_{n, \mp},
\end{align}
and 
\begin{align}
    (\mathcal{J}\mathcal{K})^2 = (\tilde{\cal J} \mathcal{K})^2 = 1,\quad (\mathcal{J}\mathcal{K}) (\tilde{\cal J} \mathcal{K}) = \left( -1 \right)^{L/2} \left( -1 \right)^{\cal F}.
        \label{eq: Z2}
\end{align}
We also introduce a translation operation $\mathcal{T}$ in the double Hilbert space by
\begin{align}
    \mathcal{T} \lambda_{n, \pm} \mathcal{T}^{-1} = \lambda_{n+1, \pm}. 
\end{align}
Then, we have
\begin{align}
    \mathcal{T}\,(\mathcal{J}\mathcal{K})\,\mathcal{T}^{-1} &= \left( -1 \right)^{L-1} (\tilde{\cal J} \mathcal{K}) = - (\tilde{\cal J} \mathcal{K}), \\ 
    \mathcal{T}\,(\tilde{\cal J} \mathcal{K})\,\mathcal{T}^{-1} &= \left( -1 \right)^{L-1} (\mathcal{J}\mathcal{K}) = - (\mathcal{J}\mathcal{K}), 
\end{align}
where we use the assumption that $L$ is even.
Using Eq.~(\ref{eq: Z2}), we have
\begin{align}
    \mathcal{T}\,(\mathcal{J}\mathcal{K})\,\mathcal{T}^{-1} &= \left( -1 \right)^{L/2-1} \left( -1 \right)^{\cal F} (\mathcal{J}\mathcal{K}), \\ 
    \mathcal{T}\,(\tilde{\cal J} \mathcal{K})\,\mathcal{T}^{-1} &= \left( -1 \right)^{L/2-1} \left( -1 \right)^{\cal F} (\tilde{\cal J} \mathcal{K}).
\end{align}
Therefore, we have a nontrivial phase $-1$ between translation symmetry $\mathcal{T}$ and KMS symmetry $\mathcal{J}\mathcal{K}$ or $\tilde{\cal J} \mathcal{K}$ for $\left( -1 \right)^{L/2-1} \left( -1 \right)^{\cal F} = -1$, equivalent to $L \equiv 0$ (mod $4$) in the physical subspace with $\left( -1 \right)^{\cal F} = +1$.
In such a case, the Lindbladian spectrum exhibits at least two-fold degeneracy, in a similar manner to Ref.~\cite{Hsieh-16}.
This can be considered as the LSM constraint arising from the combination of translation symmetry and KMS symmetry.

Next, we consider other antiunitary operations~\cite{Kawabata-22},
\begin{align}
    \mathcal{P} \lambda_{n, \pm} \mathcal{P}^{-1} = - \lambda_{n, \pm},\quad 
    \mathcal{R} \lambda_{n, \pm} \mathcal{R}^{-1} = + \lambda_{n, \pm},
\end{align}
which are explicitly constructed as
\begin{align}
    \mathcal{P} = \prod_{n=1}^{L/2} \left( \lambda_{2n-1, +} \lambda_{2n-1, -} \right) \mathcal{K},\quad
    \mathcal{R} = \prod_{n=1}^{L/2} \left( (\ii \lambda_{2n, +}) (\ii \lambda_{2n, -}) \right) \mathcal{K},
\end{align}
and satisfy
\begin{align}
    \mathcal{P}^2 = \mathcal{R}^2 = \left( - 1 \right)^{L/2},\quad
    \mathcal{P} \mathcal{R} = \left( -1 \right)^{L/2} \left( -1 \right)^{\cal F}.
\end{align}
Then, we introduce KMS symmetry with the unitary operators $\mathcal{J}\mathcal{P}$ and $\mathcal{J}\mathcal{R}$.
Here, we have
\begin{align}   
    \mathcal{T} \mathcal{J} \mathcal{T}^{-1} = \left( -1 \right)^{L/2-1} \left( -1 \right)^{\cal F} \mathcal{J},
\end{align}
and 
\begin{align}
    \mathcal{T} \mathcal{P} \mathcal{T}^{-1} 
    &= \prod_{n=1}^{L/2} \left( \lambda_{2n, +} \lambda_{2n, -} \right) \mathcal{K}
    = \left( -1 \right)^{L/2} \mathcal{R}
    = \left( -1 \right)^{L/2} \left( -1 \right)^{\cal F} \mathcal{P}, \\
    \mathcal{T} \mathcal{R} \mathcal{T}^{-1} 
    &= \left( -1 \right)^{L/2} \prod_{n=1}^{L/2} \left( \lambda_{2n-1, +} \lambda_{2n-1, -} \right) \mathcal{K}
    = \left( -1 \right)^{L/2} \mathcal{P}
    = \left( -1 \right)^{L/2} \left( -1 \right)^{\cal F} \mathcal{R}.
\end{align}
Thus, we have
\begin{align}
    \mathcal{T} \left( \mathcal{J}\mathcal{P} \right) \mathcal{T}^{-1} &= - (\mathcal{J}\mathcal{P}), \\
    \mathcal{T} \left( \mathcal{J}\mathcal{R} \right) \mathcal{T}^{-1} &= - (\mathcal{J}\mathcal{R}).
\end{align}
Therefore, we have a nontrivial phase $-1$ between translation symmetry $\mathcal{T}$ and KMS symmetry $\mathcal{J}\mathcal{P}$ or $\mathcal{J}\mathcal{R}$ for arbitrary $L$, which leads to the degeneracy of the Lindbladian spectrum as the LSM constraint.
Examples include the Majorana Hamiltonian $H = J \sum_{n=1}^{L} \ii \lambda_{n} \lambda_{n+1}$ and the two-body dissipators $L_{n} = \sqrt{\gamma} \lambda_{n} \lambda_{n+1}$ under the periodic boundary conditions (i.e., $\lambda_{n+L} = \lambda_{n}$).
Through the fermionic operator-state mapping, the Lindbladian reads
\begin{align}
    &\mathcal{L} = -\ii J \sum_{n=1}^{L} \left( \ii \lambda_{n, +} \lambda_{n+1, +} \right)
    -\ii J \sum_{n=1}^{L} \left( \ii \lambda_{n, -} \lambda_{n+1, -} \right)
    + \gamma \sum_{n=1}^{L} \left( \lambda_{n, +} \lambda_{n+1, +} \lambda_{n, -} \lambda_{n+1, -} \right) - \gamma L I,
\end{align}
where $I$ is the identity operator in the double Hilbert space.
A related model was investigated in Ref.~\cite{Shibata-19-Kitaev}.
Since this Lindbladian is invariant under translation $\mathcal{T}$ and KMS symmetry $\mathcal{JR}$, it is subject to the LSM constraint and exhibits the degeneracy.
In particular, it possesses the two-fold degenerate steady states, one of which is even under the KMS operation and the other is odd.

We also consider other antiunitary operations~\cite{Kawabata-22},
\begin{align}
    \mathcal{Q} \lambda_{n, \pm} \mathcal{Q}^{-1} = \mp \psi_{n, \pm},\quad 
    \mathcal{S} \lambda_{n, \pm} \mathcal{S}^{-1} = \pm \lambda_{n, \pm},
\end{align}
which are explicitly constructed as
\begin{align}
    \mathcal{Q} = \prod_{n=1}^{L/2} \left( \lambda_{2n-1, +} (\ii\lambda_{2n, -}) \right) \mathcal{K},\quad
    \mathcal{S} = \prod_{n=1}^{L/2} \left( (\ii \lambda_{2n, +}) \lambda_{2n-1,-} \right) \mathcal{K},
\end{align}
and satisfy
\begin{align}
    \mathcal{Q}^2 = \mathcal{S}^2 = +1,\quad
    \mathcal{Q} \mathcal{S} = \left( -1 \right)^{\cal F}.
\end{align}
Then, we introduce KMS symmetry with the unitary operators $\mathcal{J}\mathcal{Q}$ and $\mathcal{J}\mathcal{S}$.
Here, we have
\begin{align}
    \mathcal{T} \mathcal{Q} \mathcal{T}^{-1} 
    &= \prod_{n=1}^{L/2} \left( (\ii \lambda_{2n, +}) \lambda_{2n+1, -} \right) \mathcal{K}
    = \left( -1 \right)^{L/2-1} \mathcal{S}
    = \left( -1 \right)^{L/2-1} \left( -1 \right)^{\cal F} \mathcal{Q}, \\
    \mathcal{T} \mathcal{S} \mathcal{T}^{-1} 
    &= \prod_{n=1}^{L/2} \left( \lambda_{2n+1, +} (\ii\lambda_{2n, -}) \right) \mathcal{K}
    = \left( -1 \right)^{L/2-1} \mathcal{Q}
    = \left( -1 \right)^{L/2-1} \left( -1 \right)^{\cal F} \mathcal{S}.
\end{align}
Thus, we have
\begin{align}
    \mathcal{T} \left( \mathcal{J}\mathcal{Q} \right) \mathcal{T}^{-1} &= + (\mathcal{J}\mathcal{Q}), \\
    \mathcal{T} \left( \mathcal{J}\mathcal{S} \right) \mathcal{T}^{-1} &= + (\mathcal{J}\mathcal{S}).
\end{align}
In contrast to the other types of KMS symmetry, nontrivial phases do not appear between translation symmetry and KMS symmetry, which implies the absence of the LSM constraints.

It is also notable that total fermion parity $\left( -1 \right)^{\cal F}$ in Eq.~(\ref{aeq: fermion parity total}) commutes with the translation operation $\mathcal{T}$:
\begin{align}
    \mathcal{T} \left( -1 \right)^{\cal F} \mathcal{T}^{-1} = \left( -1 \right)^{\cal F}.
\end{align}
As a result, no constraints arise even in the simultaneous presence of $\left( -1 \right)^{\cal F}$ and $\mathcal{T}$.
By contrast, fermion parity $\left( -1 \right)^{F_{\pm}}$ in the individual ket or bra space anticommutes with $\mathcal{T}$:
\begin{align}
    \mathcal{T} \left( -1 \right)^{F_{\pm}} \mathcal{T}^{-1} = - \left( -1 \right)^{F_{\pm}},\quad \left( -1 \right)^{F_{\pm}} = \prod_{n=1}^{L/2} \left( \ii \lambda_{2n-1, \pm} \lambda_{2n, \pm} \right).
\end{align}
Consequently, the Lindbladian spectrum exhibits the degeneracy---LSM constraint due to the combination of strong fermion parity $\left( -1 \right)^{F_{\pm}}$ and translation $\mathcal{T}$ in open quantum systems.


\end{document}